\documentclass[10pt,twocolumn]{article}

\usepackage{amsmath}
\usepackage{cite}
\usepackage{graphicx}
\usepackage{siunitx}
\usepackage{textcomp}
\usepackage{microtype}
\usepackage{subcaption}
\usepackage{multirow}
\usepackage{authblk}
\usepackage{fancyhdr}
\usepackage{url}
\usepackage{color,soul}

\title{High-sensitivity free space optical communications using low size, weight and power hardware}

\author[1,*]{Alexander D. Griffiths}
\author[1]{Johannes Herrnsdorf}
\author[2]{Oscar Almer}
\author[2]{Robert K. Henderson}
\author[1]{Michael J. Strain}
\author[1]{Martin D. Dawson}

\affil[1]{Institute of Photonics, University of Strathclyde, Glasgow}
\affil[2]{CMOS Sensors \& Systems Group, University of Edinburgh, Edinburgh}

\affil[*]{Corresponding author: alex.griffiths@strath.ac.uk}

\begin{document}
	\twocolumn[
	\begin{@twocolumnfalse}
		\maketitle
		\begin{abstract}
			
			Free space optical communication systems with extremely high detector sensitivities are attractive for various applications with low size, weight and power requirements. For these practical systems, integrated hardware elements with small form factor are needed. Here, we demonstrate a communication link using a CMOS integrated micro-LED and array of single-photon avalanche diodes. These integrated systems provide a data rate of 100~Mb/s at a sensitivity of $-55.2$~dBm, corresponding to 7.5 detected photons per bit.
		\end{abstract}

  \end{@twocolumnfalse}
]

\section{Introduction}

	Intensity modulated optical communication has been shown to provide multi-Gb/s data rates over free space \cite{Rajbhandari2016a}, and is expected to be part of the next generation of wireless communication systems \cite{Haas2015}. Additionally, high sensitivity receivers and optimised transmission schemes enable data transmission with very low levels of received power \cite{Chitnis2015,Robinson2005,Robinson2006}. Typically, high data rate and high sensitivity experimental demonstrations are performed using large, complex or high power consumption equipment, such as arbitrary waveform generators, CW lasers with external modulators and superconducting cryogenic receivers. Further encoding and decoding complexity is introduced when high order modulation schemes and multiplexing techniques are employed \cite{Rajbhandari2016a}. These transmitter and receiver hardware requirements can be problematic for deployment in application areas where low size, weight and power (SWaP) systems are desirable.
	
	Recently, single-photon avalanche diodes (SPADs) have attracted interest for high sensitivity optical communications \cite{Zhang2018}. A SPAD is a silicon device which produces electrical pulses on detection of a single photon, followed by a "dead time", in which it is insensitive to incoming light. By fabricating arrays of SPADs and combining the outputs in either a digital or analog fashion, high dynamic range optical receivers can be produced \cite{Almer2015,Gnecchi2016a,Kosman2019}. As SPAD fabrication is compatible with current silicon complementary metal-oxide-semiconductor (CMOS) technology, highly integrated receiver systems can be developed, with significant signal processing performed on-chip \cite{Gnecchi2016,Steindl2018}. The single-photon nature of these receivers allows exceptionally high sensitivity levels to be reached, moving closer to the standard quantum limit (SQL) than more conventional avalanche photodiodes (APDs) can reach \cite{Steindl2018}. The SQL is determined by the Poissonian nature of photon detection, and gives the minimum number of photons required to achieve a given BER \cite{Zimmermann2017,ShiehOFDM}. When considering specific data rates and photon wavelengths, this gives a limit on receiver sensitivity, usually quoted in dBm. At 100~Mb/s, with 635~nm light, sensitivities as low as $-51.6$~dBm have been demonstrated, 18.5~dB from the SQL of $-70.1$~dBm \cite{Zimmermann2017}.
	
	Gallium nitride micro-LEDs have been shown to have high modulation bandwidths, enabling Gb/s data rate communications when combined with high order modulation schemes \cite{Rajbhandari2016a}. Additionally, micro-LEDs can be fabricated in high-density array format and bump-bonded to CMOS control electronics, providing compact, integrated devices with a digital interface \cite{McKendry2009}. The high degree of spatial and temporal control over the optical emission of these devices, without the need for a digital-to-analog converter, makes them attractive as transmitters for optical communications \cite{Zhang2013,Griffiths2017}. 
	
	Here, we present an optical communication link implemented with a CMOS controlled micro-LED transmitter and a SPAD array receiver. As both transmitter and receiver are realised with integrated electronic systems, the dependence on large, power-hungry hardware is lifted. With a simple transmission scheme, data rates of 50 and 100~Mb/s are demonstrated with sensitivities of $-60.5$ and $-55.2$~dBm respectively. In addition, the combined power consumption of the current, unoptimised system is less than 5.5~W, demonstrating that these compact, digitally-interfaced devices can provide high performance optical communication links on strict SWaP budgets.

\section{Methods}

	The data transmission scheme employed here is return-to-zero (RZ) on-off keying (OOK). The implementation of this scheme is simple, as the transmitter modulates between two output intensity levels, and the receiver can decode the stream with a single threshold. Despite the higher modulation bandwidth requirements, RZ transmission was chosen over non-return to zero (NRZ) as it has been show to improve bit error ratio (BER) performance in SPAD based systems by reducing inter-symbol interference (ISI) \cite{Zimmermann2017}. In NRZ transmission, the timing jitter of photon detection events can cause an overflow into the next bit period. However, in RZ transmission, there is an interval prior to transmission of each bit in which no photons are sent, reducing the probability that detection events overflow.
	
	\begin{figure}[htb]
		\centering
		\includegraphics[width=\linewidth]{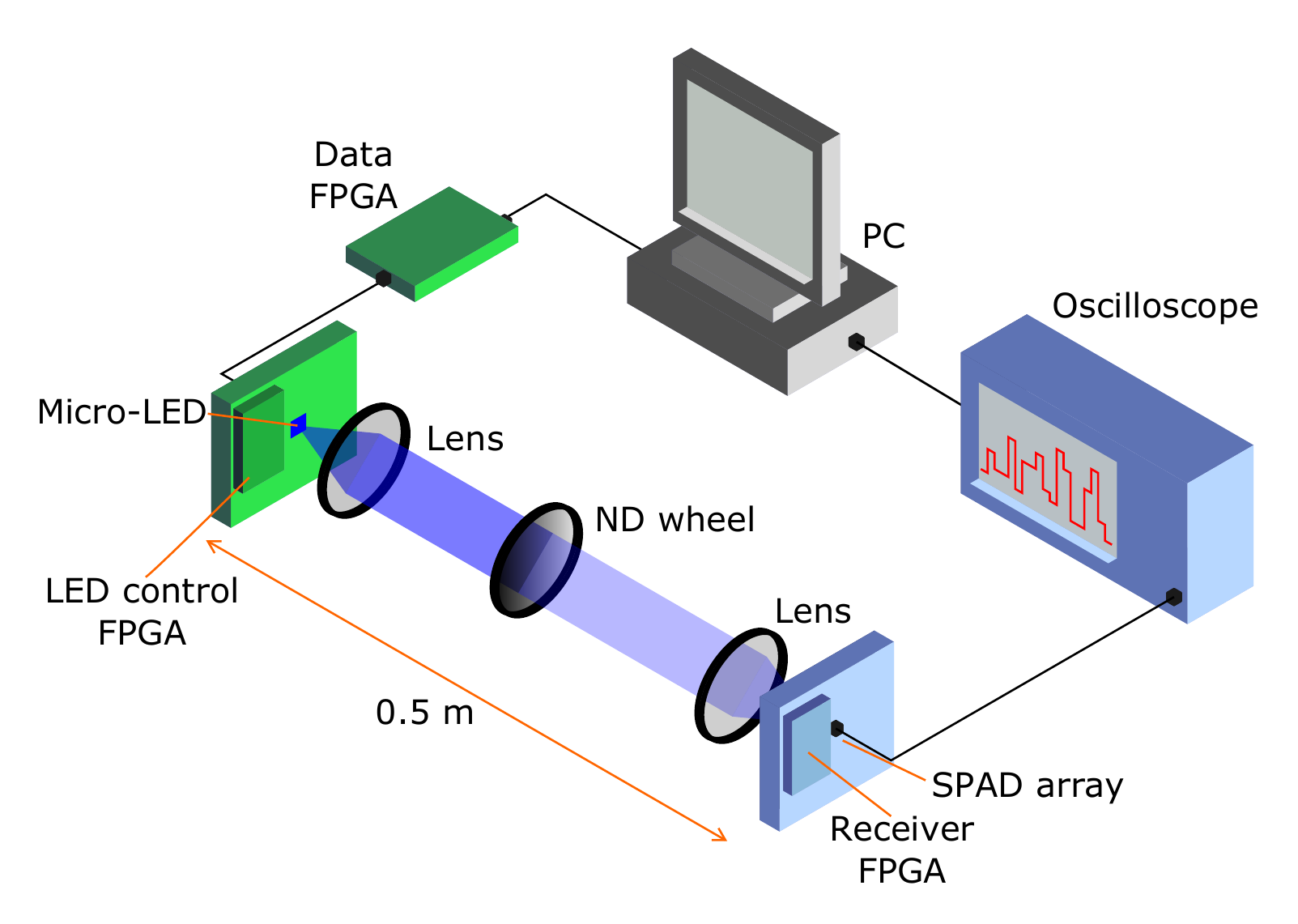}
		\caption{Schematic of the experimental setup.}
		\label{fig:Setup}
	\end{figure}
	
	The experimental arrangement is shown schematically in Figure \ref{fig:Setup}, and was operated in a dark laboratory. The optical transmitter used is a single, $99 \times 99$~\SI{}{\micro\meter\squared} pixel from a $16 \times 16$ array chip of micro-LEDs, emitting at 450~nm. The fabrication details and characterisation of similar devices are reported in reference \cite{Zhang2013}. The micro-LED array is bonded to CMOS control electronics, enabling individual control of the driving conditions for each pixel. This approach provides a mm-scale device containing optical emission elements and control electronics. The device is housed on an evaluation printed circuit board (PCB) where a field-programmable gate array (FPGA) (Opal Kelly XEM3010) provides control signals and electrical power.
	
\begin{figure}[htb]
		\centering
		\includegraphics[width=\linewidth]{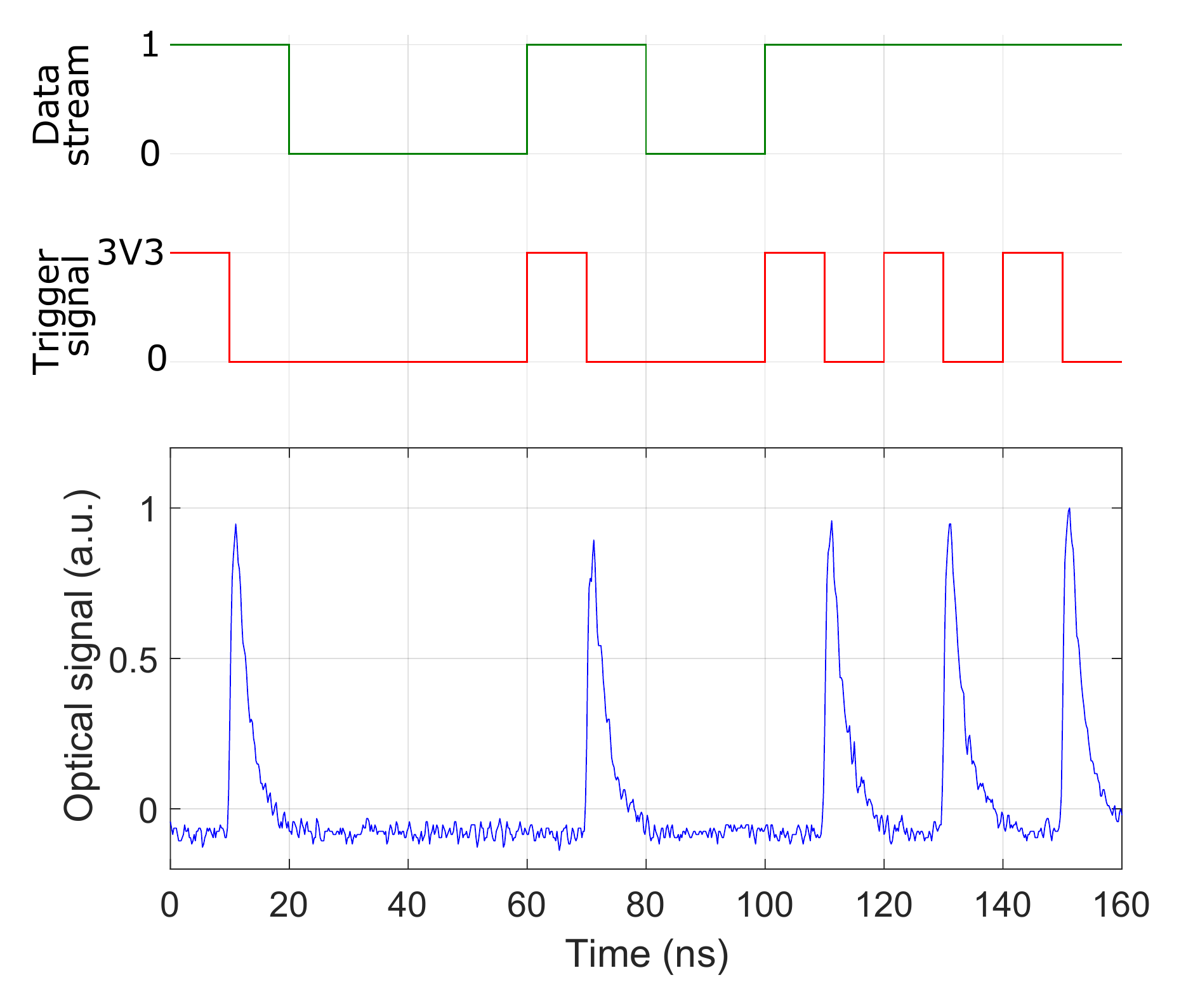}
		\caption{The 50~Mb/s binary data stream (upper), is converted to a suitable trigger signal for NRZ-OOK (middle), which produces optical emission from the LED (lower).}
		\label{fig:Waveforms}
	\end{figure}

	The CMOS drive electronics were designed with a mode that allows short pulse generation from the LEDs, where the falling edge of an input logic signal triggers a short electrical driving signal \cite{McKendry2009}. A pseudorandom bit sequence (PRBS) can be adjusted to provide this trigger signal by operating at a 50\% duty cycle. The resulting sequence is loaded on to a separate FPGA (Opal Kelly XEM6310), which provides the trigger signal to the micro-LED board at the desired data rate. The optical emission from the micro-LED, captured with an APD (Hamamatsu C5658) is shown in Figure \ref{fig:Waveforms}, along with the data stream and trigger signals. The full width at half maximum of the optical pulse is 3~ns wide. In the following experiments, data rates of 50 and 100~Mb/s are used, resulting in RZ-OOK duty cycles of 15\% and 30\%, and emitted average power of \SI{1.62}{\micro\watt} and \SI{3.56}{\micro\watt}, respectively. The higher value at 100~Mb/s is a result of the pulse frequency being doubled, and the pulses not having time to fully relax back to no emission. There is a half bit period delay introduced by the falling edge triggering mechanism, however, this is easily accounted for during the decoding process.
	
	For data transmission, the optical signal is collimated with a lens (Thorlabs C220TME-A) and transmitted over a lab bench distance of 0.5~m. A graded neutral density (ND) filter wheel (Thorlabs NDC-50C-4M-A) is used in the optical path to control the attenuation of the signal reaching the receiver. A collection lens (Thorlabs ACL4532U) focuses the light onto the active area of the optical receiver.
	
	The SPAD array receiver consists of $64 \times 64$ SPADs on a \SI{21}{\micro\meter} pitch, with a fill factor of $F_{fill} = 43\%$ \cite{Kosman2019}. Including the surrounding electronics, the chip is $2.6 \times 2.8$~\SI{}{\milli\meter\squared}, and packaged to interface with a PCB. A further FPGA (Opal Kelly XEM6310) provides control signals and power to the chip, and an external 15~V bias is applied to the SPAD pixels. The individual pixels have a photon detection probability (PDP) of $\eta_{PDP}= 26\%$ at 450~nm \cite{Richardson2009a}, a measured dark count rate of 350~Hz and a dead time of 20~ns. The output of the SPADs is combined using XOR trees and ripple counters so the device is operated as a digital silicon photomultiplier (dSiPM), with a single output of photon counts at a given sample rate \cite{Gnecchi2016,Abbas2018}. For the experiments here, a $32 \times 32$ subset of the array was set as active, in order to reduce the total dark count level while still enabling a large dynamic range. Additionally, an ND filter with 3\% transmittance was placed infront of the active area to reduce background counts from stray light around the system.
	
	The photon count signal from the chip is read out from low voltage differential signaling (LVDS) pads using a differential probe and oscilloscope. Using the FPGA interface, the array was set to output the number of received photons at a rate of 200~MHz, with a range of 0-31 photon counts. The analog voltage signal from the LVDS pads was captured by the oscilloscope, resampled and digitized to recover a photon count signal. To decode the bit stream, photon counts are integrated over the bit period and compared to a threshold value. 
	
	As the output from the receiver is a number of photons at a given sample rate, the detected photons per second ($\Phi_{det}$) is readily obtained from the oscilloscope trace. Incident photons per second ($\Phi_{inc}$) can then be calculated according to:
	\begin{equation}\label{eq:Inc_ph}
	\Phi_{inc} = \frac{\Phi_{det}}{\eta_{PDE}(1-\Phi_{det}\frac{\tau_d}{N})}.
	\end{equation}
	Here, $N$ is the number of active SPADs, $\tau_d$ is the dead time of a single SPAD, and $\eta_{PDE}$ is the photon detection efficiency of the array, given by $\eta_{PDE} = \eta_{PDP}F_{fill}$. Incident optical power $P_{inc}$ is then given by:
	\begin{equation}
	P_{inc} = \Phi_{inc}E_{ph},
	\end{equation}
	where $E_{ph}$ is the energy of a 450~nm photon.

\section{Results \& Discussion}

\begin{figure}[htb]
		\centering
		\includegraphics[width=\linewidth]{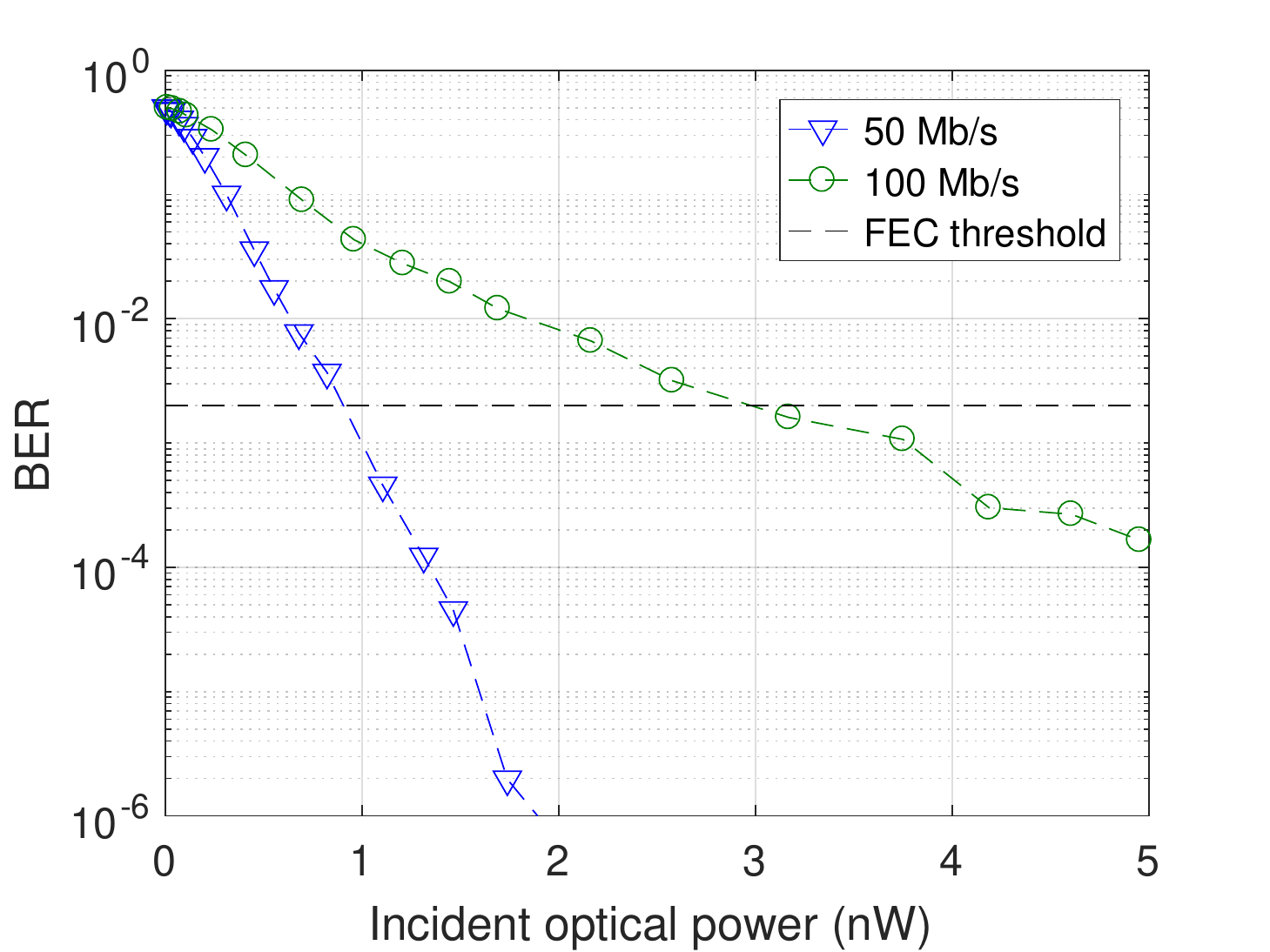}
		\caption{BER against incident optical power incident on the receiver for data rates of 50 and 100 Mb/s.}
		\label{fig:BER}
	\end{figure}

	To asses the BER performance of the communication link, the PRBS of $2^{15}$ bits was repeatedly transmitted until over $10^6$ bits were received. Experiments were performed at 50 and 100~Mb/s, for various incident power levels, producing the BER curves in Figure \ref{fig:BER}. A target BER threshold of \SI{2e-3}{} is plotted for reference, as this is sufficient for forward error correction (FEC) codes to achieve an output BER of \SI{1e-9}{} \cite{FEC}. The signal properties at the FEC threshold are summarized in Table \ref{tab:FEC_Signal}. An incident optical power of 0.9 and 3.0~nW is required for 50 and 100~Mb/s respectively, corresponding to sensitivities of $-60.5$ and $-55.2$~dBm.

	\begin{table*}[htb]
		\centering
		\caption{Received optical power ($P$) at the FEC threshold for 50 and 100~Mb/s, with corresponding sensitivity ($S$), distance to the SQL ($D$), average detected photons per bit ($\phi_{det}$) and average incident photons per bit ($\phi_{inc}$).}
		\label{tab:FEC_Signal}
		\begin{tabular}{ccccccc}\hline
			$R_{data}$	&	$P$	&	$S$		& 	$D$				&	$\phi_{det}$&$\phi_{inc}$	\\ 
			(Mb/s)		&	(nW)&	(dBm)	& 	(dB)			&	(ph/bit)	&	(ph/bit)	\\\hline
			50			&	0.9	&	-60.5	&	11.1			&	4.6			&	41			\\
			100			&	3.0	&	-55.2	&	13.42			&	7.5			&	68			\\ \hline
		\end{tabular}
	\end{table*}
	
\begin{figure}[htb]
		\centering
		\begin{subfigure}{\linewidth}
			\includegraphics[width=\linewidth]{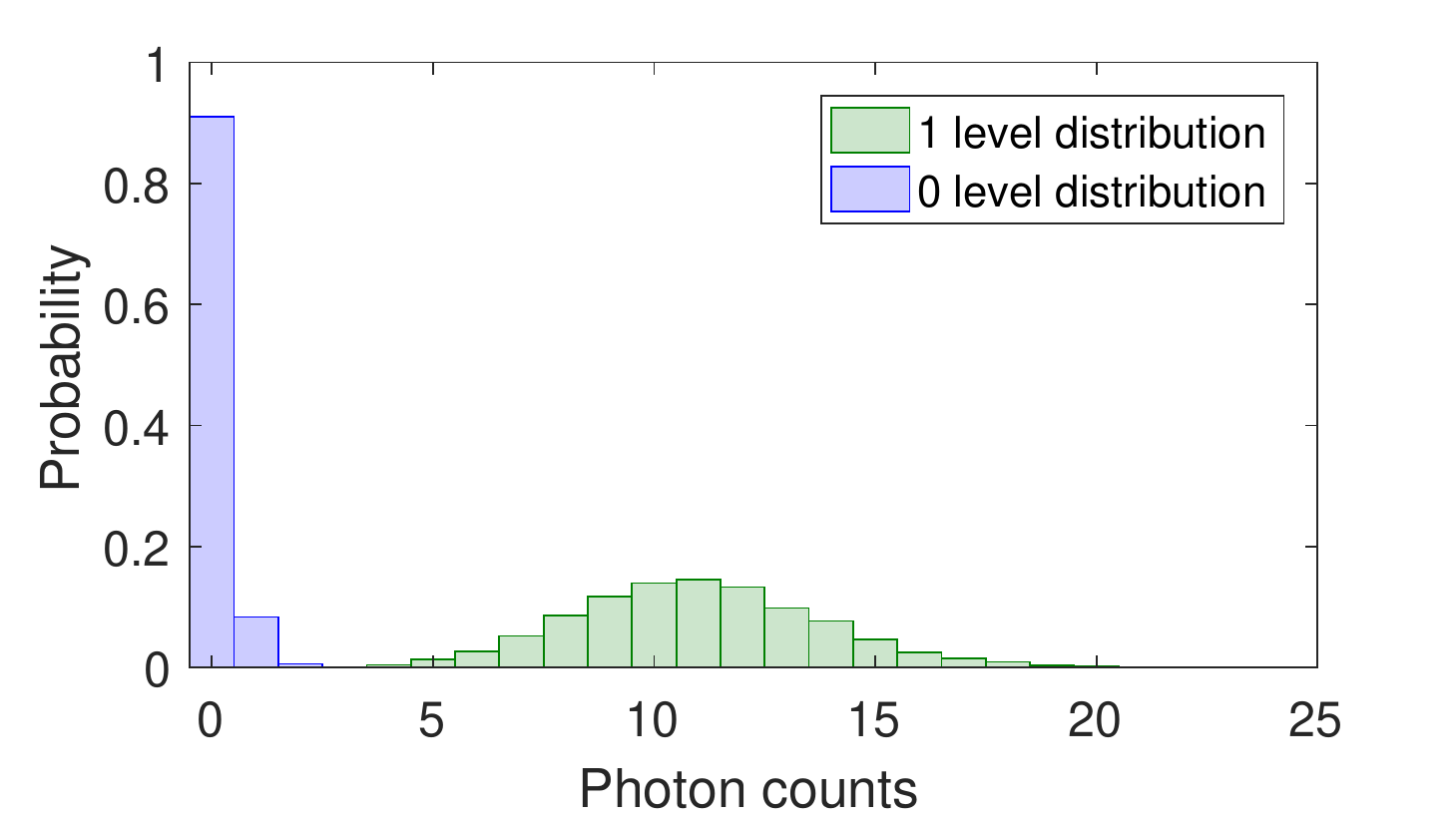}
			\caption{}
		\end{subfigure}
		~
		\begin{subfigure}{\linewidth}
			\includegraphics[width=\linewidth]{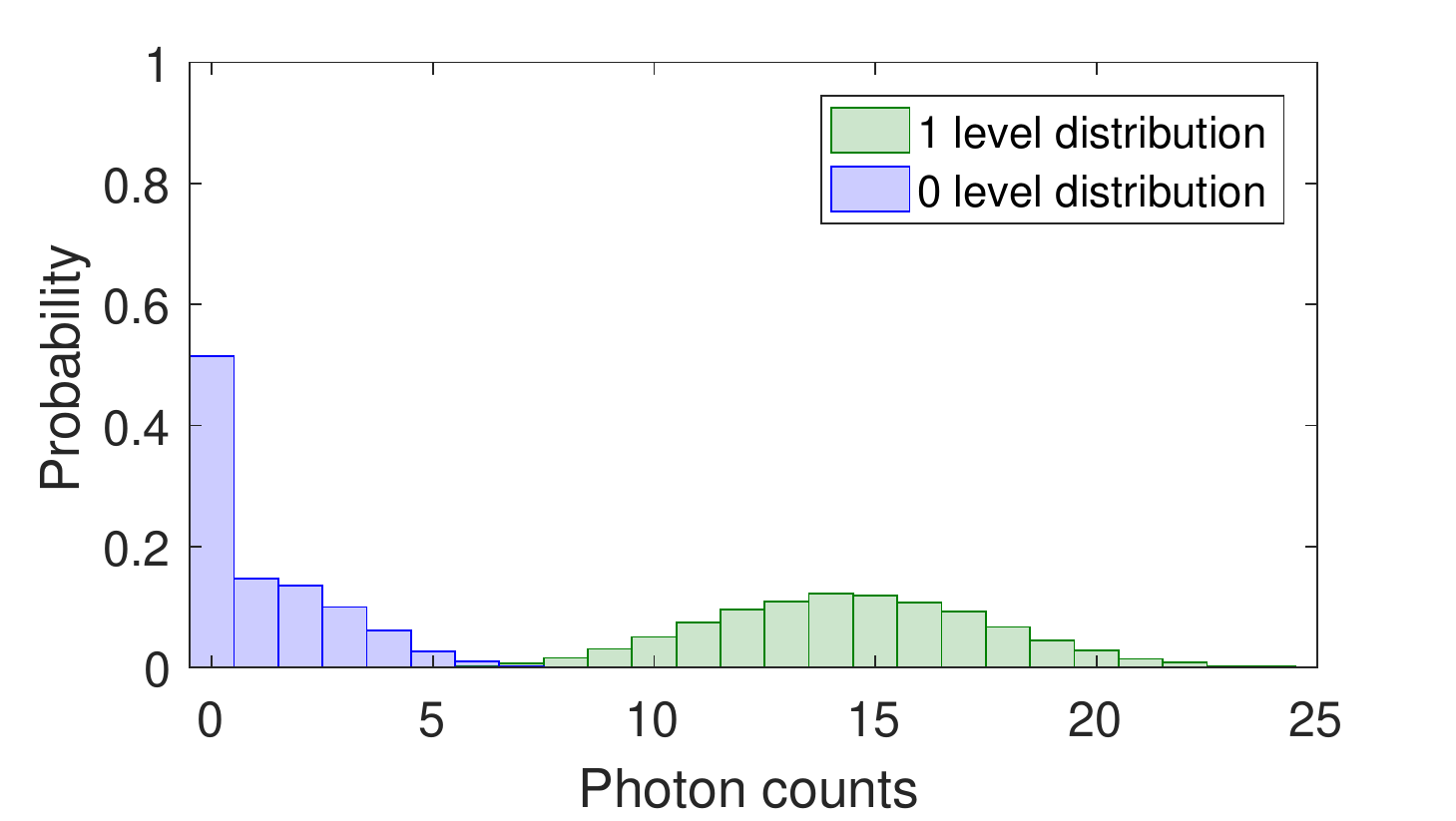}
			\caption{}
		\end{subfigure}
		\caption{Probability distributions of 0 and 1 level photon counts per bit for (a) 50 and (b) 100 Mb/s at a BER of \SI{1e-3}{}.}
		\label{fig:Distributions}
	\end{figure}

	 At 50~Mb/s with 450~nm photons and a target BER of \SI{2e-3}{}, the attained sensitivity of the system is 11.1~dB away from the SQL of $-71.6$~dBm. The 26\% PDP and 43\% fill factor accounts for 9.5~dB of the difference, with the remainder attributed to the effects of background and dark counts. This can be seen in Figure \ref{fig:Distributions}a, which shows the experimental probability distributions of photon counts for transmission of a binary ``0'' and ``1'' at the FEC threshold for 50~Mb/s. Both closely follow a Poisson distribution around their mean, and can be readily distinguished by applying a decision threshold to identify the transmitted bit. The non-zero count levels for transmission of a ``0'' push the threshold requirements to higher levels, increasing the power requirements above the SQL. Nevertheless, an average of only 4.6 detected photons is required to achieve the FEC threshold. Accounting for PDP and fill factor with Equation \ref{eq:Inc_ph}, this corresponds to 41 incident photons per bit. 
	
	For 100~Mb/s, the attained sensitivity is 13.4~dB from the SQL of $-68.6$~dBm. It would be expected to remain the same distance as for 50~Mb/s, as the photon counts per bit should remain the same, however the distance is increased due to ISI. Figure \ref{fig:Distributions}b shows the probability distribution of counts for 100~Mb/s. Here it can be seen that the ``0'' level distribution is no longer Poissonian. In fact, it is the average of two Poisson distributions, associated with whether the previous transmitted bit was ``0'' or ``1''. If the previous bit is ``1'', some photons from the LED pulse trail into the next bit, due to the longer tail of emission seen in Figure \ref{fig:Waveforms}. This pushes the decision threshold to a higher level, increasing the requirement on incident power for distinguishing a ``1'' from a ``0''. This is a limitation from the pulse performance of the micro-LED, and may be lifted with devices operating with shorter pulse widths. Despite this limitation, only 7.5 detected photons are required per bit to reach the FEC threshold, corresponding to 68 incident photons per bit.
	
	Reference \cite{Steindl2018} provides a useful review of state-of-the-art high sensitivity SPAD receivers. The work presents an integrated SPAD receiver system with sensitivities of $-51.2$ and $-46.3$~dBm for 50 and 100~Mb/s with a BER of \SI{2e-3}{}. Sensitivities of $-46.2$ and $-43.8$~dBm for 150 and 200~Mb/s respectively are shown for a higher BER of \SI{6.5e-3}{}. In both cases a 635~nm laser with external modulator was used as the optical transmitter. Our work shows a sensitivity improvement of 14.3 and 11.4~dB respectively over the lower data rates at a wavelength of 450~nm, where the SQL is also higher due to the higher photon energy. While we are currently unable to reach data rates above 100~Mb/s due to limited FPGA output rates and micro-LED pulse widths, related work with the same receiver and a laser diode transmitter shows the same trend in sensitivity enhancement \cite{Kosman2019}.
	
	Both transmitter and receiver used here are mm-size, chip-scale devices housing LEDs, drive electronics, photodetector arrays and receiver electronics. The current system houses these chips in ceramic packages connected to evaluation PCBs, which are $13 \times 18.5$~\SI{}{\centi\meter\squared} and $12.5 \times 20.5$~\SI{}{\centi\meter\squared} for transmitter and receiver respectively. While this hardware is already at a PCB level and relatively compact, many parts of the evaluation boards would be unused in a final application, allowing transceiver systems to be developed on the scale of a few square centimeters. 
	
	The electrical power for the system is drawn by four elements: the data FPGA, LED control FPGA, receiver FPGA and receiver bias. Power for the micro-LED emitter and CMOS control electronics is drawn through the LED control FPGA. The SPAD photodetectors are powered through the additional receiver bias, while the surrounding electronics draw power through the receiver FPGA. The power consumption of the system is summarized in Table \ref{tab:Power}, with the total transmitter and receiver combined consuming 5.48~W. The current arrangement has not been optimised for power consumption, so this can be considered an upper ceiling on the requirements. Power requirements can be readily reduced by streamlining the FPGA configurations, combining the data and control boards, and moving to application-specific integrated circuits (ASICs). Estimating from the CW performance of similar devices in reference \cite{Zhang2013}, the micro-LED pixels show a low-current, wall-plug efficiency of around 1-2\%. With an emitted optical power of \SI{1.62}{\micro\watt}, this indicates that the power consumption of the LED control FPGA is not dominated by the optical emitter. Additionally, the SPAD receiver consumes 115~mW when considered without the FPGA \cite{Kosman2019}. Furthermore, many application areas may already be employing FPGAs or on-board computer systems which could be used to produce and process the digital signals, meaning only the LED and receiver hardware consume additional power.
	
	\begin{table}[htb]
		\centering
		\caption{Power consumption of components in the system, totalling 5.48~W.}
		\label{tab:Power}
		\begin{tabular}{cc}\hline
			Component			&	Power draw (W)	\\ \hline
			Data FPGA			&		0.95 	\\
			LED control FPGA	&		1.00	\\
			Receiver FPGA		&		3.45	\\
			Receiver bias		&		0.08	\\ \hline
		\end{tabular}
	\end{table}

\section{Conclusion}

In conclusion, a communication link with data rates at 100~Mb/s has been demonstrated with a sensitivity of $-55.2$~dBm, 13.42~dB from the SQL. This link was achieved using a highly integrated transmitter and receiver with low form factor control electronics. While the decoding was performed offline, utilising a high-speed oscilloscope, the digital nature of the system will enable full decoding using digital electronics in the future. Additionally, the receiver can operate at higher output frequencies and count ranges, however, the signal becomes increasingly analog, and the decoding method could not reliably convert the oscilloscope voltage trace back to integer photon counts. A future system which directly acquires a digital signal through the receiver FPGA will bypass the need for this conversion and greatly increase the dynamic range of the receiver. Furthermore, the CMOS controlled micro-LED arrays have been demonstrated for multi-level communications, which could exploit this enhanced dynamic range to enable higher data rates with only minor adjustments to FPGA configurations \cite{Griffiths2017}. 

The small form factor, lack of complex hardware, simple transmission scheme and low power consumption enables high data rate, high sensitivity data communications on a low SWaP budget. An optimised version of the hardware used here could be employed in low SWaP applications, without strongly affecting payload capabilities or power systems.
\\ \\
The underlying data for this work is available at: \url{http://dx.doi.org/10.15129/fe315481-2a4a-4b84-a8ea-b239ddf27074}

\section*{Funding Information}

This work was funded by the UK Engineering and Physical Sciences Research Council (EPSRC) (EP/M01326X/1, EP/S001751/1)

% Bibliography
\small
\bibliography{Bibliography}
\bibliographystyle{ieeetran}

\end{document}